 \documentclass [12pt,a4paper      ]{article}
\usepackage{graphics}
\usepackage{times}

\DeclareFontFamily{OT1}{times}{}
\DeclareFontShape {OT1}{times}{m }{n }{ <-> ptmr }{}
\DeclareFontShape {OT1}{times}{bx}{n }{ <-> ptmb }{}
\DeclareFontShape {OT1}{times}{m }{it}{ <-> ptmri}{}
\DeclareFontShape {OT1}{times}{bx}{it}{ <-> ptmbi}{}
\usepackage{amsmath}
\usepackage{amsfonts}
\usepackage{amssymb}
\usepackage{latexsym}
\newcommand{\cl}{C \kern -0.1em \ell} 
\newcommand{\bbR}{\mathbb{R}}         
\newcommand{\bbC}{\mathbb{C}}         
\newcommand{\bbH}{\mathbb{H}}         
\newcommand{\bbB}{\mathbb{B}}         

\newcommand{\DEF}{:=}                 

\newcommand{\VEC}{\vec{\kern +.1em[}} 
\newcommand{\TOR}{\vec{\kern +.2em]}} 
\newcommand{\BRA}{\langle\kern -.2em\langle} 
\newcommand{\KET}{\rangle\kern -.2em\rangle} 

\begin{document}

\title{\bf\vspace{-2.5cm} Quaternions in mathematical physics (1):\\
                             \emph{Alphabetical bibliography}\footnote{Living report, to be updated by the authors, first posted on ~{\tt arXiv.org}~ on October 16, 2005, anniversary day of the discovery of quaternions, on the occasion of the bicentenary of the birth of William Rowan Hamilton (1805--2005).} }

\author{Andre Gsponer and Jean-Pierre Hurni\\ ~\\ 
\emph{Independent Scientific Research Institute}\\ 
\emph{Oxford, OX4 4YS, England}}

\renewcommand{\today}{6 July 2008}
\date{ISRI-05-04.26 ~~ \today}

\maketitle

\begin{abstract}

This is part one of a series of four methodological papers on (bi)quaternions and their use in theoretical and mathematical physics: 1~- Alphabetical bibliography, 2~- Analytical bibliography, 3~- Notations and definitions, and 4~- Formulas and methods.  

This quaternion bibliography will be further updated and corrected if necessary by the authors, who welcome any comment and reference that is not contained within the list.

\end{abstract}

\begin{figure}
\begin{center}
\resizebox{6cm}{!}{ \includegraphics{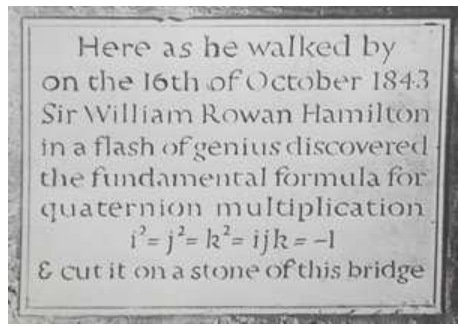}}
\end{center}
\caption{\emph{Plaque to William Rowan Hamilton at Brougham Bridge on the Royal Canal, now in the Dublin suburbs. \copyright 2004, Royal Irish Academy.}}
\label{fig:1}
\end{figure}

\section{\Huge Introduction}
\label{int:0}

The two component formulation of complex numbers, and the non-commutative algebra of quaternions, are possibly the two most important discoveries of Hamilton in mathematics.  Using modern vector algebra notation, the quaternion product can be written \cite{HAMIL1891-}
\begin{equation}\label{int:1}
                              [a+\vec{A}] [b+\vec{B}]
                            = ab-\vec{A}\cdot\vec{B} 
                            + a\vec{B} + b\vec{A} + \vec{A}\times\vec{B} ,
\end{equation}
where $[a+\vec{A}]$ and $[b+\vec{B}]$ are two quaternions, that is four-number combinations $Q\DEF s+\vec{v}$ of a scalar $s$ and a three-component vector $\vec{v}$, which may be real (i.e, quaternions, $Q \in \bbH$) or complex (i.e., biquaternions, $Q \in \bbB$).  Comparing with the product of two complex numbers, called `biscalars' by Hamilton, i.e.,
\begin{equation}\label{int:2}
                              (a+ia')(b+ib')=ab-a'b'+iab'+iba',
\end{equation}
where $a, a'$ and $b, b'$ are pairs of real numbers, one can interpret the quaternion product \eqref{int:1} as a generalization of the complex number product \eqref{int:2}.

The reasons why Hamilton's quaternions have never become a prevalent formalism in physics, while Hamilton's formulation of complex numbers has been universally adopted, is an open question.  The fact is that ``quaternion structures'' are very frequent in numerous areas of physics, the most prominent examples being special relativity (i.e., Lorentz transformations), electrodynamics, and spin.  For this reason quaternions and their generalizations keep reappearing in a number of forms, which are as numerous as diverse: spinors, Einstein's semivectors, Pauli matrices, Stokes parameters, Eddington numbers, Clifford numbers, qubits, etc.  The same is true in mathematics, where because of the isomorphisms $SU(2) \sim \bbH / \bbR$,\,  $SL(2,\bbC) \sim \bbB^\star{\! /} \bbC$,\,  $\bbB = \bbC \otimes \bbH \sim M_2(\bbC) \sim \cl_{3,0}$,\,  $\bbH \otimes \bbH \sim \cl_{0,3}$, etc., ``quaternion structures'' also arise in many different contexts.

On the other hand, because quaternion algebra yields more efficient algorithms than matrix algebra for three- and four-dimensional applications, their use in computer simulations and graphics, numerical and symbolic calculations, robotics, navigation, etc., has  become more and more frequent in the past few decades.  It is therefore not surprising that some of the largest published quaternion bibliographies contain many references to these and other practical applications \cite{ELL---2005-}.

\emph{The purpose of the present bibliography is to present a selection, but as comprehensive as possible, of the use of biquaternions in theoretical and mathematical physics, with an emphasis on their applications to fundamental rather than applied topics.}\footnote{The only exceptions to this rule are papers or books of general interest, and papers included for completeness when they are useful to understand other papers.}   

  However, the bibliography is not restricted to just papers in which quaternions or biquaternions are used explicitly.  Quite the contrary, the bibliography also covers papers in which a hypercomplex whole-symbol system allied to quaternions is used (e.g., Clifford-numbers, Pauli vector-matrices, Eddington-numbers, semivectors, two-component spinors, twistors, etc.), and papers in which a quaternion or biquaternion structure plays a central role.  The kind of papers that are not included are those in which a strictly conventional matrix-type formalism is used (e.g., the Pauli- or Dirac-matrices, and the corresponding standard two- or four-component vector formalisms), and papers which have not been published (or would not qualify to appear) in a peer-reviewed journal.

  A final important criterion used in compiling the bibliography is that it includes only papers that we have read, so that we were able to attach a few keywords to each entry in the reference list.  These keywords have the format \%\%KEYWORD, where ``\%'' is the symbol used for comments in \TeX, so that they  do not appear in the compiled bibliography.  However, the keywords are visible and can be searched for in the  \TeX source of the bibliography (see next section).  Moreover, the keywords can be used to sort and filter the bibliography according to diverse criteria.  This will be done for instance in the second paper in this series \cite{GSPON2005E}, where the keywords will be defined and used to produce an ``analytical'' version of the ``alphabetical'' bibliography listed in the present paper.

  The bibliography has grown from 230 entries in the 1993 version, i.e., Ref.\,\cite{GSPON1993A}, to 1430 entries in the present version.

  Because they have often been considered by authors working with quaternions, a separate section at the end of the quaternion bibliography (i.e., Sec.\,\ref{qbib}) is dedicated to octonions (i.e., Sec.\,\ref{obib}).

\newpage

\section{\Huge Format of typical references}
\label{for:0}

What follows is a list of typical references in which the label, keywords, and directives (to be explained in paper two of this series, Ref.~ \cite{GSPON2005E}) are made visible.

\begin{itemize}

\item {\bf Book:}

{\bf LANCZ1949-} C. Lanczos, The Variational Principles of Mechanics (Dover, New-York, 1949, 1986) 418~pp.  Quaternions pages 303--314. \%\%BOOK, \%\%QUATERNION, \%\%SPECIAL-RELATIVITY, \%\$D06022002.

\item {\bf Conference proceedings, festschrift or contributed volume:}

{\bf SPROS1996-} W. Spr\"ossig and K. G\"urlebeck, eds., Proc. of the Symp. ``Analytical and Numerical Methods in Quaternionic and Clifford Analysis,'' Seiffen, June 5--7, 1996 (TU Bergakademie Freiberg, 1996) 228~pp.  \%\%BOOK, \%\%QUATERNION, \%\%ANALYTICITY, \%\%CLIFFORD, \%\$D20032002.

\item {\bf Paper in a conference proceedings, festschrift, or contributed volume:}

{\bf PENRO1990-} R. Penrose, \emph{Twistors, particles, strings and links}, in: D.G. Quillen et al., eds., The Interface of Mathematics and Particle Physics (Clarendon Press, Oxford, 1990) 49--58. \%\%QUATERNION, \%\%TWISTOR, \%\$D05022002.

\item {\bf Paper in a journal:}

{\bf WEISS1941-} P. Weiss, \emph{On some applications of quaternions to restricted relativity and classical radiation theory}, Proc. Roy. Irish Acad. {\bf A 46} (1941) 129--168. \%\%QUATERNION, \%\%MAXWELL, \%\%SPINOR, \%\%LORENTZ-DIRAC, \%\$D09022002.

\item {\bf Two authors:}

{\bf EINST1932-} A. Einstein and W. Mayer, \emph{Semi-Vektoren und Spinoren}, Sitzber. Preuss. Akad. Wiss. Physik.-Math. Kl. (1932) 522--550. \%\%QUATERNION, \%\%SEMIVECTOR, \%\$D05022002.

\item {\bf Report or preprint:}

{\bf VELTM1997-} M. Veltman, \emph{Reflexions on the Higgs system}, Report 97-05 (CERN, 1997) 63~pp. \%\%BOOK, \%\%QUATERNION, \%\%LEPTODYNAMICS, \%\%GAUGE-THEORY,  \%\$D14052001, \%\$C Veltman uses quaternions in the form of 2x2 matrices.

\item {\bf Ph.D. thesis:}

{\bf GURSE1950A} F. G\"ursey, \emph{Applications of quaternions to field equations}, Ph.D.\ thesis (University of London, 1950) 204~pp.  \%\%BOOK, \%\%QUATERNION, ~~\%\%DIRAC, ~~\%\%GENERAL-RELATIVITY, ~~\%\%LANCZOS, \%\%PROCA, \%\$D20032002. 

\item {\bf Unpublished document}:

{\bf GSPON1993A} A. Gsponer and J.-P. Hurni,  \emph{Quaternion bibliography: 1843--1993}. Report ISRI-93-13 (17 June 1993) 35~pp. This is the first version of the present bibliography, with 228 entries. \%\%QUATERNION, \%\%BIBLIOGRAPHY, \%\$D20032002.

\item {\bf ``arXived'' e-print:}

{\bf GSPON2002D} A. Gsponer, \emph{Explicit closed-form parametrization of $SU(3)$ and $SU(4)$ in terms of complex quaternions and elementary functions}, Report ISRI-02-05 (22 November 2002) 17~pp.;  e-print \underline{ arXiv:math-ph/0211056 }.  \%\%QUATERNION, \%\%ALGEBRA, \%\%HADRODYNAMICS, \%\$10092005.

\item {\bf Internet page:}

{\bf NOBIL2006-} R. Nobili, \emph{Fourteen steps into quantum mechanics}, HTML document (Posted in 2006) about 13 pp.;  available at\\ \underline{ http://www.pd.infn.it/~rnobili/qm/14steps/14steps.htm }. \%\%QUATERNION, \%\%QUANTUM-PHYSICS, \%\%QQM, \%\$D25.

\end{itemize}

\newpage

\section{\Huge Quaternion bibliography}
\label{qbib}

\newpage

\section{\Huge Octonion bibliography}
\label{obib}

\section*{\Huge ACKNOWLEDGMENTS}
\label{ACKNOWLEDGMENTS}

This bibliography would not exist without the help, dedication, and professionalism of Mrs.\ Claire-Lise Held and Mrs.\ Jocelyne Favre at the University of Geneva Physics department library, and many other librarians at other universities.

\end{document}